%% file: main.tex
\documentclass[10pt,conference]{IEEEtran}
\usepackage[utf8]{inputenc}
\usepackage[T1]{fontenc}    
\usepackage{hyperref}       
\usepackage{url}            
\usepackage{booktabs}       
\usepackage{amsfonts}       
\usepackage{nicefrac}       
\usepackage{microtype}      
\usepackage{comment}

\usepackage{graphicx}
\usepackage{doi}
\usepackage{amsfonts}
\usepackage{xcolor}
\usepackage{physics}
\usepackage{makecell}
\usepackage{multirow}
\usepackage{graphicx}
\usepackage{subcaption}
\usepackage{enumitem}
\usepackage{duckuments}

\usepackage{tikz}
\usetikzlibrary{positioning}

\usepackage{physics}
\usepackage{algorithm}
\usepackage{algpseudocode}

\usepackage{mathtools} 
\usepackage{array}
\usepackage{wrapfig} 
\usepackage{svg}
\usepackage[capitalise]{cleveref}
\usepackage[natbib=true, sorting=none, maxcitenames=1, style=numeric-comp]{biblatex}

\usepackage[colorinlistoftodos,prependcaption,textsize=tiny]{todonotes}

\usepackage{subfiles}

\definecolor{white}{RGB}{255, 255, 255}
\definecolor{black}{RGB}{0, 0, 0}
\definecolor{iceblue}{RGB}{166, 187, 200}
\definecolor{orange}{RGB}{245, 130, 32}
\definecolor{green}{RGB}{23, 156, 125}
\definecolor{navy}{RGB}{0, 91, 127}
\definecolor{bluegreen}{RGB}{0, 133, 152}
\definecolor{turquois}{RGB}{57, 193, 205}
\definecolor{lime}{RGB}{178, 210, 53}
\definecolor{aquadark}{RGB}{51, 124, 153}
\definecolor{aqualight}{RGB}{102, 157, 178}
\definecolor{skydark}{RGB}{153, 189, 204}
\definecolor{skymedium}{RGB}{204, 222, 229}
\definecolor{skylight}{RGB}{229, 238, 242}
\definecolor{vwblue}{RGB}{28, 63, 82}
\definecolor{beige}{RGB}{211, 199, 174}
\definecolor{gold}{RGB}{253, 185, 19}
\definecolor{plumdark}{RGB}{124, 21, 77}
\definecolor{plumlight}{RGB}{187, 0, 86}
\usepackage{lineno}
\begin{document}

\title{Is data-efficient learning feasible with quantum models?}

\author{Alona Sakhnenko\IEEEauthorrefmark{1}\IEEEauthorrefmark{2}, Christian B. Mendl\IEEEauthorrefmark{2}, Jeanette M. Lorenz\IEEEauthorrefmark{1}\IEEEauthorrefmark{3} \\
	\IEEEauthorblockA{\IEEEauthorrefmark{1}Fraunhofer Institute for Cognitive Systems IKS,  Munich, Germany}
    \IEEEauthorblockA{\IEEEauthorrefmark{2}Technical University of Munich, Munich, Germany}
	\IEEEauthorblockA{\IEEEauthorrefmark{3}Ludwig-Maximilian University, Munich, Germany}
}
\maketitle
\begin{abstract}
The importance of analyzing nontrivial datasets when testing quantum machine learning (QML) models is becoming increasingly prominent in literature, yet a cohesive framework for understanding dataset characteristics remains elusive. In this work, we introduce a data-generation tool that allows to construct semi-artificial classical datasets tailored to quantum kernel methods (QKMs). Using this tool, we show that on fully classical datasets, QKMs can require fewer training examples than classical kernels to reach comparable error, providing
clear empirical evidence that data-efficient learning with quantum models is possible on classical data.  The main motivation behind this tool is to enable the community to perform controlled studies to figure out which dataset characteristics are particularly fitting for quantum models by tuning the data-generation procedure. Additionally, our study brings a spectral-bias–based generalization metric from classical kernel methods into the QML domain and show that the performance predicted by this metric aligns closely with empirical results, thereby closing an important gap between theory and practice in QML generalization. Our tool paves the way for a systematic exploration of dataset complexities. This could potentially contribute to a deeper understanding of the generalization benefits of QKM models (extendable to a broader family of QML models) and shifts the search for quantum advantage from ad hoc benchmark hunting to principled dataset design.

\end{abstract}

\section{Introduction}

\subfile{sections/1_introduction.tex}

\subfile{sections/figure_1.tex}

\section{Background}\label{sec:background}
\subsection{Generalization in QML}\label{sec:generalization}
\subfile{sections/2_generalization.tex}

\subsection{Inductive (spectral) bias}\label{sec:alternative}
\subfile{sections/3_method.tex}

\section{Results}\label{sec:results}
\subfile{sections/5_results.tex}

\section{Discussion}\label{sec:discussion}
\subfile{sections/6_discussion.tex}

\section{Conclusion}

In this work,  we provided empirical evidence that there exist fully classical datasets for which quantum kernel methods (QKMs) can learn more data-efficiently than standard classical kernels. To enable a study of such instances accessible in a controlled way, we introduce a semi-artificial data–generation tool to generate datasets tailored specifically to a chosen QKM  and illustrate a concrete motivating use case for its application. This work facilitates the development of probing techniques aimed at identifying specific use cases that can effectively leverage quantum advantages in contrast to the currently dominant practice of testing quantum machine learning (QML) models on generic datasets in a hope to find one where QMLs show practical advantage. 

We achieve this by employing the target-alignment metric proposed by \cite{Canatar_2021} to generate semi-artificial datasets where QKMs exhibit greater data efficiency than classical Kernel Machines (KMs). Across our experiments, the underlying spectral‑bias generalization metric that forms the basis of this target‑alignment shows strong predictive performance in the QKM setting, adding another valuable tool to the QML practitioner's toolkit. 

Overall, our work contributes to a deeper understanding of QKM's potential for data efficiency and lays a solid foundation for future studies in dataset characterization for potential quantum advantage as well as extending these insights to broader classes of QML architectures.

\section*{Acknowledgements}
The research is part of the Munich Quantum Valley, which is supported by the Bavarian state government with funds from the Hightech Agenda Bayern Plus.

\printbibliography

\begin{appendix}
\subsection{Saturating geometric difference}\label{sec:geometric_diff}
For completeness, we compare our method with an established method that was proposed by \citet{Huang_2021} and investigated in our previous work \cite{Egginger_2024}. The aim of this method is to maximize a geometric difference between quantum $K_Q$ and classical $K_C$ kernel matrices. This is done by defining a matrix $\sqrt{K_Q}(K_C + \lambda Id)^{-1}\sqrt{K_Q}$, where $\lambda = 1.1$ is a regularization term. We then select eigenvector $\mathbf{v}$ that corresponds to the largest absolute eigenvalue. The new labels are assigned as $\tilde{y} = \sqrt{K_Q}\mathbf{v}$. 

\begin{figure}
    \includegraphics[width=0.45\textwidth]{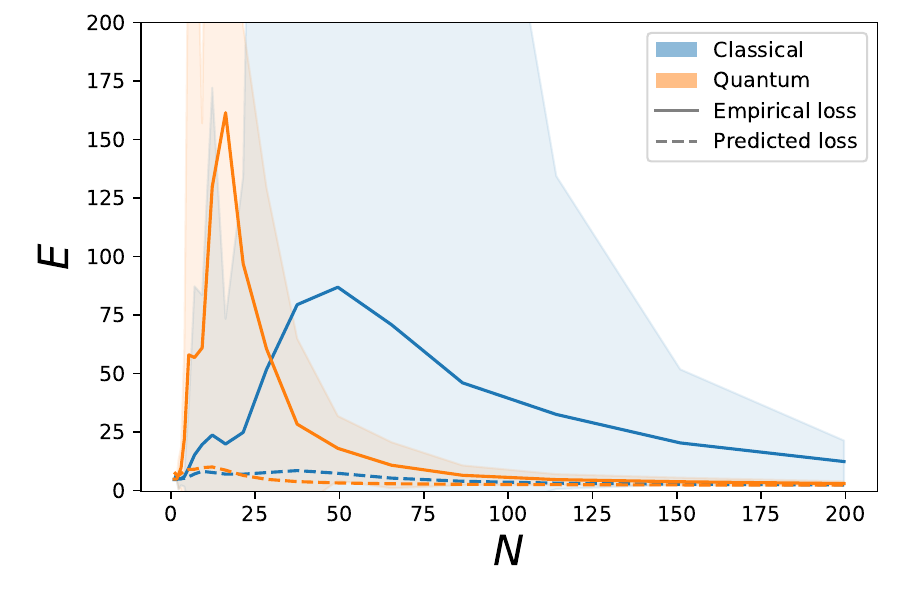}
\caption{Square mean error loss $E$ difference between classical and quantum kernels that depends on $N$ on dataset relabeled to saturate geometric difference}
\label{fig:geom_results}
\end{figure}

As shown in \cref{fig:geom_results}, a similar trend in performance gaps between quantum and classical kernels is observed with the dataset aimed at saturating the geometric difference between these kernels; however, this gap is less pronounced as in the method we proposed as the difference between the quantum and classical kernels is negligible, especially for lower number of points. Interestingly, the predicted loss on this dataset shows a notable divergence from the empirical loss. The underlying reasons for this discrepancy warrant further investigation in future research.

\subsection{Bandwidth effect}\label{sec:bandwidth}

The theoretically justified explanation provided by \cite{canatar2023bandwidthenablesgeneralizationquantum} outlines how bandwidth facilitates generalization in quantum contexts, supporting its significance as a hyperparameter. In an empirical architecture study conducted by \cite{Egginger_2024}, bandwidth emerged as the most important hyperparameter influencing model performance. In the main experiment, bandwidth was set to $t=0.5$. Here, we explore whether changing the bandwidth to $t=1.0$ (effectively removing it) has any significant impact on the performance. 

\begin{figure}
    \centering
    \includegraphics[width=0.45\textwidth]{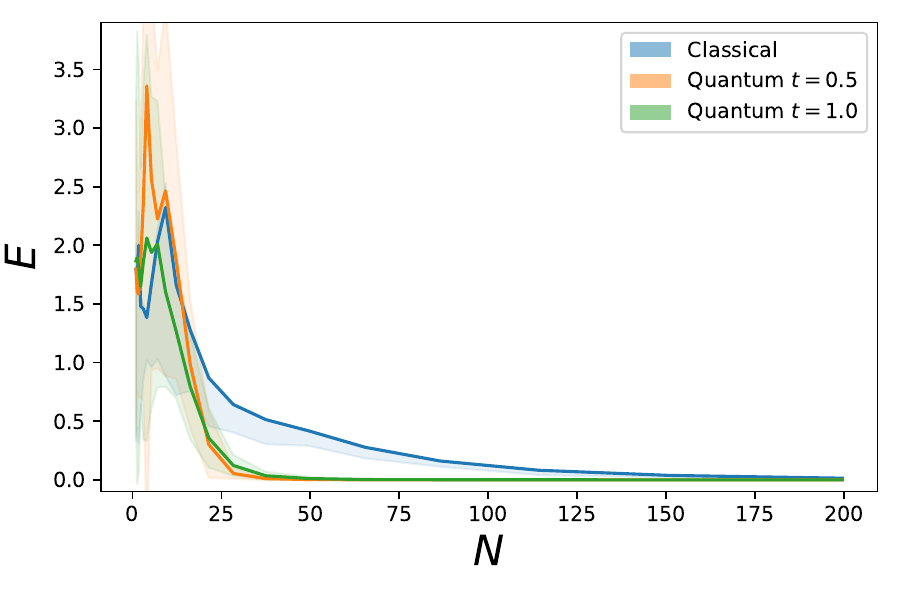}
    \caption{Effects of bandwidth $t$ hyperparameter on convergence of error rate tested on semi-artificial dataset with $\hat{c}^{20}_1$.}
    \label{fig:bandwidth}
\end{figure}

As illustrated in \cref{fig:bandwidth}, the impact of bandwidth on overall performance appears to be minimal, with no significant effects observed on error rates or convergence rates. Notably, at 
$t = 1.0$, the error rate for smaller datasets $N$ is lower, which contradicts the findings of our previous study \cite{Egginger_2024}, which indicated a preference for smaller values of $t$ in QKMs. To validate these findings, a more extensive study is warranted.

\end{appendix}

\end{document}

%% file: sections/1_introduction.tex
Aside from the speed up opportunities offered by larger quantum computing systems, a potential advantage of quantum machine learning models lies in their generalization benefits over classical models. These efforts often aim to be dataset-agnostic, but this frequently  results in unpredictable empirical performance. In light of the "no free lunch" theorem, which asserts that no single model excels across all datasets, it is sensible to focus more on the qualities of the datasets in conjunction with the capabilities of the models. Recent literature \cite{bermejo2024quantumconvolutionalneuralnetworks, ablan2025similaritybandwidthtunedquantumkernels} emphasizes the significance of exploring \textit{nontrivial datasets} for accessing advantage with QML models, which hints at the importance of shifting away from the dataset-agnostic approaches commonly used in the field.

This leads to an important questions: \textit{(i) what are the nontrivial datasets? (ii) how to characterize the datasets  that are trivial for QML models but hard for classical learners}? Within the QML community, various approaches have been explored to identify which dataset properties can lead to an advantage when utilizing QML models~\cite{schetakis2021binaryclassifiersnoisydatasets, Huang_2021, Meyer_2023,  Glick_2024}. However, these ideas remain fragmented and lack a unifying framework. One recurring observation across various studies, including those in classical machine learning, is that the size of the dataset plays a significant role in determining its complexity. Meaning, that larger datasets tend to provide models with more informative patterns, and therefore improving their performance and generalization ability, while smaller (yet complex) datasets remain challenging. This raises a question: \textit{is it feasible for a QML model be more data-efficient than their classical counterparts and if so, when?}

To address the question, selecting the right analytical tool is key. Making sweeping statements about all QML models is unrealistic, but examining a broad family of models can yield valuable insights. Kernel Methods (KMs) provide a solid theoretical basis for understanding generalization and recent work established a connection~\cite{neural_kernel, Schuld_supervised} between KMs and Deep Learning Models (DLMs), indicating that insights from KMs could have far-reaching implications. A study on the generalization capabilities of KMs by \citet{Canatar_2021} provided robust empirical support for the theoretically predicted loss observed in both KMs and DLMs, reinforcing the theoretical groundwork established for KMs. However, a similar notion is still lacking in the QML context.

In this work, we focus on a question if and when QML models can be more data-efficient than their classical counterparts (require less data for training) through the lens of QKM models. 
The contribution of this work is threefold: 
\begin{enumerate}
    \item \textit{Data Generation Tool for Exploration of Dataset Complexities:} We introduce a data generation method that produces semi-artificial datasets, which are better aligned with a given QKM. This tool enables controlled construction and systematic exploration of classical datasets on which QKMs are expected to perform well.
    \item \textit{Empirical Evidence of Data-Efficient Learning on Classical Data:} Using this tool on several real-world tabular datasets, we show that QKMs can achieve comparable generalization error with fewer training examples than standard classical kernels, thereby demonstrating that quantum models can be more data-efficient even when trained on fully classical data.
    \item \textit{Empirical Test of Spectral-Bias–Based Generalization Metric for QML:} We empirically show that spectral bias generalization metric \citet{Canatar_2021} (illustrated in \cref{fig:generalization-metrics}) reliably predicts empirical performance of a QKM on a given dataset, in contrast to other tested metrics that were reported to lack such predictive power in prior work.
\end{enumerate}

This paper is structured in the following way. We begin with reviewing important background notions (\cref{sec:background}), where we review existing generalization metrics and their empirical limitations (\cref{sec:generalization}), introduce the alternative metric of \citet{Canatar_2021} as a promising tool for QML (\cref{sec:alternative}),  and discuss potential dataset complexity measures to quantify it (\cref{sec:complexity}). We then use the alternative metric to formulate a way to generate a semi-classical, which are designed is way to allow quantum KMs to achieve low error rates with less data points (\cref{sec:method}). We show the effectiveness of our technique and hence provide an evidence of existence of dataset on which this is possible as well as test the generalization metric (\cref{sec:results}) and we discuss the implication of our findings on a broader family of models in (\cref{sec:discussion}).

%% file: sections/figure_1.tex
\begin{figure}
\begin{subfigure}[t]{0.45\textwidth}
    \centering
\begin{tikzpicture}
    \tikzset{
        node/.style={rectangle,very thick,draw=black,text=black, minimum size=5mm, align=center, minimum width=2.5cm},
        visible/.style={rectangle,very thick,draw=gold,text=black, minimum size=5mm, minimum width=2.5cm, align=center},
        textnode/.style={text=black},
        textnodeblack/.style={text=black}
    }
    \node[node] (root) {\textbf{QML theory}};
    \node[node] (1_l) [below left = 0.25cm and -1cm of root]{Challenges};
    \node[node] (1_r) [below right = 0.25cm and -1cm of root]{Advantages};
    
    \node[node] (2_ll) [below left =0.25cm and -1.25cm of 1_l]{Trainability};
    \node[textnode](path_2_ll) [below =-0.15cm of 2_ll] {\footnotesize\cite{McClean_2018, thanasilp2024exponentialconcentrationquantumkernel}};
    \node[node] (2_lr) [below left= 1.5cm and -2.5cm of 1_l]{Non simulatability};
    \node[textnode](path_2_lr) [below =-0.15cm of 2_lr] {\footnotesize\cite{bermejo2024quantumconvolutionalneuralnetworks, ablan2025similaritybandwidthtunedquantumkernels}};
    \node[textnode](path_2_l) [above left =0.08cm and -1.5cm of 2_lr] {\footnotesize\cite{ablan2025similaritybandwidthtunedquantumkernels}};

    \node[visible] (2_rl) [below right= 1.5cm and -2.5cm of 1_r]{Resource efficiency};
    \node[node] (2_rr) [below right = 0.25cm and -1.25cm of 1_r]{Generalization};
    \node[textnode](path_2_r) [above right=0.08cm and -1.65cm of 2_rl] {\footnotesize\cite{Caro_2022, Huang_2021}};

    \node[visible] (complexity) [below = 4cm of root]{$\mathcal{X}$ complexity};
    \node[visible] (size) [below right = 0.45cm and -2.25cm of 2_rl]{$\mathcal{X}$ size};
    \node[textnode](path_2_l) [below =-0.15cm of size] {\footnotesize\cite{Huang_2021}};

    \node[node] (4_l) [below left = 0.25cm and -0.75cm of complexity]{Biased solution};
    \node[textnode](path_2_l) [below =-0.15cm of 4_l] {\footnotesize\cite{Liu_2021, Glick_2024}};
    \node[visible] (4_r) [below right = 0.25cm and -0.75cm of complexity]{General solution};
    \node[textnode](path_2_l) [below =-0.15cm of 4_r] {\footnotesize\cite{Abbas_2021}};

    \node[node] (solution) [below = 6.25cm of root]{\textbf{QML architectures}};
     \node[textnodeblack](path_4) [above =0.1cm of solution] {\footnotesize\cite{kubler2021inductive, bowles2023contextualityinductivebiasquantum}};
    
    \path[draw=black, ->, shorten <=2pt, shorten >=2pt]
        (root) edge[bend right] node[left =0.01cm] {} (1_l)
        (root) edge[bend left] node[right =0.01cm] {} (1_r)
        (1_r) edge node[left =0.01cm] {} (2_rr)
        (1_r) edge[bend right] node[left =0.01cm] {} (2_rl)
        (1_l) edge node[left =0.01cm] {} (2_ll)
        (1_l) edge[bend left] node[left =0.01cm] {} (2_lr)
        (4_l) edge[bend right] node[left=0.01cm] {} (complexity)
        (solution) edge[bend left] node[left=0.01cm] {} (4_l)
        (solution) edge[bend right] node[left=0.1cm] {} (4_r)
    ;

    \path[draw=gold, ->, shorten <=2pt, shorten >=2pt]
        (4_r) edge[bend left] node[left=0.01cm] {} (complexity)
        (4_r) edge[bend right] node {} (2_rr)
    ;
    
    \path[draw=black, <->, shorten <=3pt, shorten >=3pt, dashed]
        (4_l) edge[bend right] node {} (4_r)
        (complexity) edge[bend right] node {} (size)
    ;

    \path[draw=black, <->, shorten <=3pt, shorten >=3pt]
        (2_ll) edge node {} (2_lr)
    ;
    
     \path[draw=gold, <->, shorten <=3pt, shorten >=3pt]
        (2_rl) edge node {} (2_rr)
        (2_lr) edge node {} (complexity)
        (2_rl) edge node {} (complexity)
        (2_rl) edge node {} (size)
    ;

    \end{tikzpicture}
    \caption{Conceptual landscape of QML theory. Yellow indicates concepts that are relevant for this work.}
    \label{fig:lit_overview}
    \end{subfigure}
    ~
    \begin{subfigure}[t]{0.45\textwidth}
        \begin{tikzpicture}
    \draw[thick,dotted,rounded corners=5, fill=red!50, fill opacity=0](0,2.5)rectangle ++(3.8,5.25)node[midway]{};  
    \draw[thick,dotted,rounded corners=5, fill=yellow!50, fill opacity=0] (4.1,2.5)rectangle ++(3.8,5.25)node[midway]{};  

    \draw[thick,dotted,rounded corners=5, draw=black] (4.5,2.65) rectangle ++(3,2.75)node[midway]{};  

    \draw[thick,dotted,rounded corners=5, fill=blue!50, fill opacity=0] (-0.25,4.5)rectangle ++(8.4,3)node{}; 
    \draw[thick,dotted,rounded corners=5, fill=orange!50, fill opacity=0] (-0.25,3)rectangle ++(8.4,1)node[midway]{};  

    \node[draw, draw opacity=0] at (2,8.3) {\textbf{Kernel methods}};
    \node[draw, draw opacity=0] at (2,7.9) {\textit{Theoretically understood}};
    \node[draw, draw opacity=0] at (6.25,8.3) {\textbf{Neural networks}};
    \node[draw, draw opacity=0] at (6.25,7.9) {\textit{Favourable scalability}};

    \node[draw, draw opacity=0, rotate=-90, color=black] at (7.7,4) {\textbf{Overparametrized}};

    \node[draw, draw opacity=0, rotate=90] at (-0.4,6) {\textbf{Quantum}};
    \node[draw, draw opacity=0, rotate=90] at (-0.4,3.5) {\textbf{Classical}};

    \draw[thick, draw=bluegreen] (0.1,6.55) rectangle ++(3.5,0.5)node[midway]{\small Geometric difference \cite{Huang_2021}};
    \draw[thick, draw=gold] (0.1,5.8) rectangle ++(3.5,0.5)node[midway]{\small Inductive bias \cite{kubler2021inductive}};

    \draw[thick, draw=bluegreen] (4.25,6.8) rectangle ++(3.5,0.5)node[midway]{\small Covering numbers \cite{Caro_2022}};
    \draw[thick, draw=lime] (4.25,6.15) rectangle ++(3.5,0.5)node[midway]{\small Fourier spectrum \cite{Schuld_2021}};
    \draw[thick, draw=lime] (4.25,5.5) rectangle ++(3.5,0.5)node[midway]{\small Effective dimensions \cite{Abbas_2021}};

    \draw[thick, draw=gold, fill=gold!5] (0.6,3.25) rectangle ++(6.8,0.5)node[midway]{\small Spectral bias \cite{Canatar_2021}};

    \draw[thick, dashed, fill=gold!5, draw=gold] (0.6,4.7) rectangle ++(6.8,0.5)node[midway]{};
    \draw[thick, dashed, fill=gold!10, draw=gold!50] (0.8,4.75) rectangle ++(2.5,0.4)node[midway]{\small This work};
    
    \draw[fill=gold, draw=gold] (0.5,2)rectangle ++(0.5,0.15)node[midway]{}; 
    \node[draw, draw opacity=0] at (1.4, 2.075) {Bias};

    \draw[fill=bluegreen, draw=bluegreen] (2.3,2)rectangle ++(0.5,0.15)node[midway]{}; 
    \node[draw, draw opacity=0, align=left] at (3.8, 2.05) {Complexity};

    \draw[fill=lime, draw=lime] (5.1, 2)rectangle ++(0.5,0.15)node[midway]{}; 
    \node[draw, draw opacity=0, align=left] at (6.6, 2.05) {Expressivity};

    \draw [-stealth,thick, draw=gold](3,3.75) -- (3,4.7);
    \draw [-stealth,thick, draw=gold](5.75,3.75) -- (5.75,4.7);
    \end{tikzpicture}
    \caption{Generalization metrics zoo for classical and quantum KMs and NNs.}
    \label{fig:generalization-metrics}
    \end{subfigure}
        \caption{Comprehensive overview of the interconnections among key concepts: This graph illustrates the relationships and interplay between various ideas, highlighting how they influence and inform one another in the broader context.}
\end{figure}

%% file: sections/2_generalization.tex
Generalization in (Q)ML refers to a model's capability to perform well on data beyond the finite training dataset. One way to estimate generalization is through empirical validation, typically done by evaluating the model's performance on a separate validation dataset. However, this approach only provides insights based on a limited number of model instances and datasets. To deepen the understanding of generalization, researchers often aim to establish theoretical upper bounds on generalization error, giving a broader view of a model's performance across various scenarios. This theoretical framework usually involves accepting specific assumptions (as shown in \cref{fig:generalization-metrics}), which we will discuss in detail below.

A common assumption is the \textit{complexity assumption}, which presumes that simpler models are less prone to overfitting compared to more complex ones and, hence, generalize better. However, this hypothesis has been recently called into question in the context of an overparametrized regime~\cite{Zhang2021, Gil_Fuster_2024}, where large models (with substantial number of trainable parameters) can still generalize well despite their complexity. Nevertheless, metrics based on this assumption remain a popular choice in the field. For example, a covering numbers complexity metric has been utilized to assess generalization error of quantum~\cite{Caro_2022} and hybrid~\cite{wu2025generalizationboundshybridquantumclassical} convolutional NNs (CNNs). This assessment enabled \citet{Caro_2022} to demonstrate that the amount of data required for training quantum CNNs is lower than previously anticipated. 
In context of KMs, \citet{Huang_2021} provided a formulation of one the the first theoretically grounded upper bounds on the generalization gap between quantum and classical learners.
An intriguing insight from this work is the impact of data quantity on the classical learnability of quantum model outputs. Specifically, as the amount of data available to the classical learner increases, the likelihood of successfully learning the quantum model's input-output mapping also rises. 

A related notion to the one above is the \textit{expressibility assumption}, which presumes that models that are capable to express larger families of target function are more versatile. This idea has been examined through various approaches. \citet{Abbas_2021} explored the effective dimensions of quantum versus classical neural networks, which captures geometric properties of the model space. This framework was originally developed for classical NN and adapted for the quantum domain by the authors.  Their experiment showed that a class of quantum models have higher expressivity than their classical counterparts on a tested use-cases, opening exciting new research avenues. Another method by \citet{Schuld_2021} approached the question of experessibility of QNNs by analyzing the frequencies that can be represented by a given quantum circuit through Fourier analysis, inspiring a wide field of further research \cite{Schreiber_2023}.  However, the expressivity of a model can negatively impact its trainability. Some studies have already explored the interplay between these two concepts \cite{PRXQuantum.3.010313, Barthe_2024}, indicating the presence of a narrow utility bottleneck.

An essential requirement for generalization metrics to be applicable for real-world use-cases is their ability to predict actual empirical performance. However, previous studies \cite{Egginger_2024, monnet2024understandingeffectsdataencoding} did not reveal a strong correlation between generalization metrics \cite{Huang_2021, Abbas_2021} and actual performance of the model, emphasizing the need for further research in this area. Therefore, in this work, we demonstrate the utility of an alternative type of generalization that utilizes model's bias.

Another promising area of research that has unfortunately not been explored as extensively empirically is the investigation of the \textit{inductive bias} in QML models. This metric captures the bias (a set of prior assumptions) that predisposes the model to learn certain datasets easier, e.g. how the structure of convolutional NNs is particularly well suited for image data. Although identifying these biases can be difficult, kernel methods can serve as a powerful tool to aid in this endeavor. The spectral properties of the kernel matrix can reveal significant insights into the generalization capabilities of the kernel method. One of the initial studies exploring the inductive bias in quantum kernel methods (QKMs) was conducted by \cite{kubler2021inductive}. The authors analyzed the maximal eigenvalues of the spectrum to identify the subspace of the reproducing kernel Hilbert space that is difficult to simulate classically, thus highlighting areas with the greatest potential for quantum advantage. While the maximal eigenvalue is a useful metric, it is insufficient as a standalone measure. Therefore, this work focuses on analyzing the entire spectrum and its alignment with the target function, as proposed in \cite{Canatar_2021}. The method from \cite{Canatar_2021} have been used to highlight the theoretical importance of one of the hyperparameters - bandwidth - for generalization of QKMs \cite{canatar2023bandwidthenablesgeneralizationquantum}, however, a full empirical study of the validity of this metric for QML setting is still outstanding.

%

%% file: sections/3_method.tex
Due to lacking empirical performance of generalization metrics based on complexity or expressivity assumption in previous studies, in this work, we focus on inductive bias based methods. For that we consider kernel ridge regression, a ML method that is theoretically well-understood compared to NNs. 

\subsubsection{Kernel ridge regression}

A \textit{kernel function} is a similarity measure between ${\mathbf{x}, \mathbf{x}'} \in \mathcal{X} \subset \mathbb{R}^D$, which corresponds (through Mercer's theorem) to an inner product of feature vectors $\psi(x)$ in the reproducing kernel Hilbert space (RKHS) $\mathcal{H}$:
\begin{equation}
    k(x, x') = \langle \psi(x), \psi(x') \rangle_{\mathcal{H}},
\end{equation}
where $\psi: \mathcal{X} \to \mathcal{H}$ is a feature map. Hence we can define an integral kernel operator $K: L_2 (\mathcal{X}) \to L_2 (\mathcal{X})$, which is a continuous analog of the Gram matrix $\mathbf{K}$ as:
\begin{equation}
    (Kf)(x) = \int k(x, x') f(x') p(x') dx',
\end{equation}
where $f: \mathcal{X} \to \mathbb{R}$ represents functions that belong to $\mathcal{H}$ and $p: \mathcal{X} \to \mathbb{R}$ is a probability density function of data $\mathcal{X}$. According to Mercer\'s theorem there exist a spectral decomposition of $K$ into eigenvalues $\gamma_i$ and eigenfunctions $\phi_i$ 
\begin{equation}
    k(x, x') = \sum_i \gamma_i \phi_i(x) \phi_i(x').
\end{equation}
Alternatively, \cite{Canatar_2021}
\begin{equation}
    \int k(x, x') \phi_{i}(x') p(x') dx' = \gamma_i \phi_i(x),\hspace{15pt} i = 1, \dots N.
\end{equation}

Kernel ridge regression is an optimization over function space
\begin{equation}
    f^* = \arg \min_{f \in \mathcal{H}}  \frac{1}{2} \sum_i^N (f(x_i) - y_i)^2 + \frac{\lambda}{2}||f||^2_{\mathcal{H}},
\end{equation}
where $N$ is a number of data samples and $\lambda$ is a ridge parameter. 

\subsubsection{Generalization}
Generalization error $E$ was defined by \citet{Canatar_2021} as follows:
\begin{equation}
    E = \biggl \langle \int p(x)(f^{*}(x) - \hat{f}(x))^2 dx \biggr \rangle_{\mathcal{X}}.
\end{equation}

Given that any target function $\hat{f} \in L_2(\mathcal{X})$ within RKHS \footnote{The eigenfunctions $\phi_k(x)$ for which $\gamma_k = 0$ are outside RKHS. \cite{kubler2021inductive} argue that functions outside of RKHS cannot be learned, but \cite{Canatar_2021} covers those functions as well in their supplementary material.} can be represented as $\hat{f}(x) = \sum_k \hat{\omega}_k \phi_k (x)$ with $\hat{\omega}_k = \langle \hat{f}, \phi_k \rangle_{\mathcal{H}}$, the generalization error $E$ can be computed as a sum of its modes $E_i$ in the following way:
\begin{align}\label{eq:generalization}
    & E = \sum_i \gamma_i \hat{\alpha}^2_i E_i \\
    & E_i = \frac{1}{\hat{\alpha}^2_i} \langle (\alpha_i^* - \hat{\alpha}_i)^2 \rangle_{\mathcal{X}} = \frac{1}{1 - b} \frac{\gamma_i}{(a + N \gamma_i)^2}\\
    & a = \lambda + \sum_i \frac{a \gamma_i}{a + N \gamma_i};  b = \sum \frac{N \gamma_i}{(a + N \gamma_i)^2}.
\end{align}
This formulation of $E$ depends on properties of the kernel matrix such as eigenvalues $\gamma_i$, data set size $N$ and complexity of the target function $\alpha_i$. The normalized mode errors are ordered according to their eigenvalues for all $N$
\begin{equation}
    \gamma_i > \gamma_i' \implies E_i < E_i'
\end{equation}
which implies that modes $i$ with large eigenvalues $\gamma_i$ are learned more rapidly as $N$ increases than modes with small eigenvalues. Consequently, kernel eigenfunctions with large eigenvalues can be estimated with kernel regression using a smaller number of samples \cite{Canatar_2021}. From here the autors \cite{Canatar_2021} formulated the task-model alignment measure as cumulatie distributions $C(i)$
\begin{equation}\label{eq:target-alignment}
    C(i) = \frac{\sum_{i'\leq i} \gamma_{i'} \hat{\alpha}^2_{i'}}{\sum_{i'} \gamma_{i'} \hat{\alpha}^2_{i'}}
\end{equation}
This implies that if the principal components of a kernel $k$ are aligned with a target function $\hat{f}$, it can be easily learned by the model \cite{kubler2021inductive}. 

\subsection{Quantum kernel methods}
By associating the quantum Hilbert space with a feature space $\mathcal{H}$, we can define a feature map that gives rise to a quantum kernel. An inner product of two quantum states defines a \textit{quantum fidelity kernel} as
\begin{equation}\label{eq:fidelity_kernel}
    k(x, x') = |\braket{\phi(x)}{\phi(x')}|^2.
\end{equation}
Quantum kernel methods embed data into high-dimensional Hilbert space \cite{Schuld_supervised}, which is prohibitively big even on current relatively small scale quantum devices for classical computers to process \cite{Arute_2019}. 

For a quantum algorithm to operate on a classical data, the data has to be embedded into a quantum state $\phi: \mathbf{x}_i \to \ket{\mathbf{x}_i}$, which exists in a complex Hilbert space $\mathcal{H}$ that is $\mathbb{C}^{2^n}$ with $n$ being the number of qubits. The significance of the choice of embedding operation for the performance of quantum models has been highlighted in the literature \cite{Abbas_2021, Schuld_2021}. In this work, we concentrate on the Hamiltonian evolution feature map as it has been proven to be powerful in QKM setting \cite{canatar2023bandwidthenablesgeneralizationquantum, Huang_2021, bandwidth}. This feature map has been developed for many-body problems, and it requires $n +1$ qubits to embed $n$ features and is defined as follows
\begin{equation}\label{eq:embedding}
    \ket{\mathbf{x}_i} = \bigg(\prod_{j=1}^n \exp \Big(-i \frac{t}{T}x_{ij} H_j^{XYZ} \Big)\bigg)^T \bigotimes_{j=1}^{n+1} \ket{\psi_j},
\end{equation}
where $H_j^{XYZ} = (X_j X_{j+1} + Y_j Y_{j+1} + Z_j Z_{j+1})$ with $X_j, Y_j, Z_j$ being the Pauli operators acting on $j$-th qubit, $\ket{\psi_j}$ is a Haar-random state, and $t$ and $T$ are hyperparameters that denote the total evolution time and number of Trotter steps, respectively. Some papers \cite{canatar2023bandwidthenablesgeneralizationquantum} \cite{Hubregtsen_2022} have applied a similar concept to \cref{eq:target-alignment}, namely kernel-target-alignment, to tweak an embedding layer to the target function. This approach shows promise, however, it is not aimed to reducing the data hunger of QKM models.

\subsection{$\mathcal{X}$ complexity}\label{sec:complexity}

Understanding which datasets are hard for classical machine learning methods but easy for quantum methods is not a straightforward task. Some work has already explored this question from ease of representation perspective. For example, \cite{Meyer_2023} investigated symmetric datasets, while \cite{Glick_2024} considered speedups on DLOG. Although these studies report impressive results, they focus on highly specialized datasets that may not proof relevant for practical applications. In contrast, \citet{Huang_2021} investigated how the size of the dataset (not its structure) affects the learnability of a QKM’s output by classical kernels. Specifically, they showed that the more data is available to classical learners, the easier it becomes for them to approximate the exact input–output mapping of a QKM, hence questioning their utility in large data setting. In this work, we take a step towards unifying these perspectives by developing a systematic way to generate a broad variety of datasets with different structural properties, with the goal of identifying which dataset characteristics are best suited to data-efficient learning with QKMs.

To illustrate how these characteristics can be studied in practice, we consider two metrics for smoothness of a function. The smoothness (i.e., the absence of abrupt changes) is fundamental to the generalization and robustness of ML models, since it curbs reactions to tiny input perturbations, reducing brittle behavior and enhancing generalization \cite{pmlr-v137-rosca20a, liu2022learningsmoothneuralfunctions}. One way to capture the smoothness is through so called Lipschitz constant. A function $f$ is called Lipschitz smooth if there exists a (Lipschitz) constant $L$ such that \cite{pmlr-v137-rosca20a}
\begin{equation}\label{eq:lipschitz}
    || f(x) - f(y) || \leq L || x - y ||,
\end{equation} which captures bounds the steepest rate of change. Lipschitz smoothness is a central tool for boosting performance \cite{10.1007/s10994-020-05929-w}, generalization \cite{liu2022learningsmoothneuralfunctions}, adversarial robustness and formal verification \cite{10.5555/3327757.3327761}, which proved useful even within QML context \cite{10821283}. Another way to capture smoothness is to consider the derivatives of different orders as well, which can be done through e.g. Sobolev $k$-norm as follows

\begin{equation}\label{eq:sobolev}
    ||f||^2_{H^k} = \sum_k \int (f^k)^2,
\end{equation}
where $H^k$ is a Hilbert space. With $k=2$ this norm captures e.g. the steepness (1st derivative) and the curvature (2nd derivative) of the function. This metric has inspired Sobolev loss and is gaining popularity with ML community \cite{yang2026deeperwiderperspectiveoptimal, Kilicsoy_2024}.

\section{Method}\label{sec:method}
To assess the potential resource efficiency benefits of a quantum model on a classical dataset, we propose to generate artificial labels for classical datasets (relabeling) that are more advantageous for quantum learners. Since only labels are being altered but features remain the same, we call these datasets semi-artificial in this work. We create new artificial labels using methods outlined in \cref{sec:artificial_labels}. These labels then substitute the true labels while features are remained unchanged, generating a semi-artificial datasets. This approach allows us to assess whether the performance disparity between quantum and classical models can be (artificially) made significant. If so, it opens opportunities to analyze the properties of such datasets and discuss their potential real-world origins. We then train and test the models on modified datasets and compute their empirical accuracy.  Additionally, we compute the predicted accuracy and compare it with the empirical results, thereby verifying the accuracy of the theoretical generalization predictions as outlined in \cref{eq:generalization}.

\subsection{Artificial labels}\label{sec:artificial_labels}

\begin{figure}
    \includegraphics[width=0.45\textwidth]{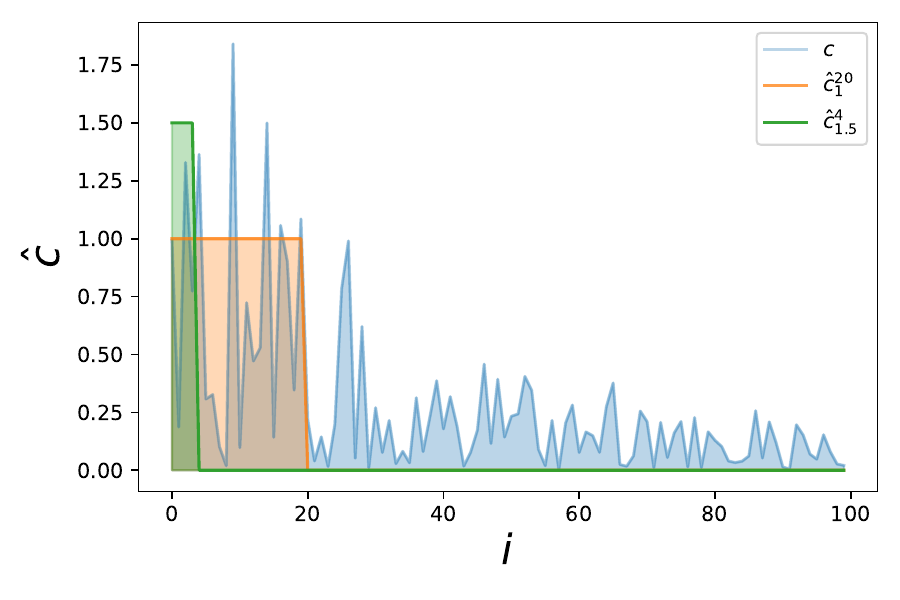}
\label{fig:hyperparametrs}
\caption{The difference between unchanged $c$ and $\hat{c}$ (\cref{eq:step_function}) with different hyperparameter settings.}
\end{figure}

Our method seeks to create an artificial target function that saturate the target-alignment measure $C(i)$ from \cref{eq:target-alignment} for quantum kernel method. This measure is a cumulative distribution captures compatibility of a chosen kernel to the target function. A steeper ascent in this cumulative distribution signifies that the function learns the target distribution more rapidly~\cite{Canatar_2021}. For computational convenience, we can express  \cref{eq:target-alignment} in matrix form with a helper vector $c = V_{\gamma}^T y ^2 = \gamma^T \hat{\alpha}^2$, each entry of the vector represents the product of the eigenvalues of the target function and the eigenvalues of the QKM. We can then represent \cref{eq:target-alignment} as follows $C(i) = \frac{\sum_{i'\leq i} c_{i'}}{\sum_{i'} c_{i'}}$. This means that for $C(i)$ to raise rapidly, we need to have $c$ in early coefficients to be high. To satisfy that we create an artificial $\hat{c}$, which we define as a step-function:
\begin{equation}\label{eq:step_function}
    \hat{c}^{n}_x(i) = 
    \begin{cases}
    x & \quad \text{if } i \leq n\\ 
    0 & \quad \text{else}, 
    \end{cases}
\end{equation}
where $n$ and $v$ are tunable hyperparameters. The difference between unaltered $c$ and relabeled $\hat{c}$ is illustrated in \cref{fig:hyperparametrs}. The new labels are assigned as
\begin{equation}\label{eq:relabeling}
    \tilde{y} = V_{\gamma}^{-T}\sqrt{\hat{c}}.
\end{equation}

\subsection{Experimental setup}
In this work, we concentrate on hybrid kernel methods by combining classical SVMs from \texttt{sklearn} library with quantum kernels implemented with \texttt{PennyLane} framework on \texttt{default.tensor} simulator. To derive the quantum kernel, we first embed the classical data points using the Hamiltonian evolution feature map due to its proven effectiveness in the QKM setting \cite{canatar2023bandwidthenablesgeneralizationquantum, Huang_2021, bandwidth} with $t=0.5$. For the classical kernel, we chose the widely used Radial Basis Function (RBF) kernel, which is a common choice in the field \cite{Huang_2021, ablan2025similaritybandwidthtunedquantumkernels, Egginger_2024}. 
We utilize the same three datasets as those in a study by \citet{moradi2022clinical}, including the Breast Cancer Wisconsin (Diagnostic), Heart Failure Clinical Records, and Pediatric Bone Marrow Transplant datasets \cite{ucimlrepo}. These datasets are particularly relevant as they present more complexity than simpler datasets like MNIST. We apply Principal Component Analysis (PCA), reducing the datasets to 8 key features. We vary the dataset size by randomly selecting rows and columns corresponding to a tested $N$ of a kernel matrix after embedding the entire dataset and repeating the experiment 50 times. To apply the methods described in \cref{sec:alternative} to these classification datasets we utilize the method proposed in \cite{Canatar_2021} and define a vector function that encodes labels in one-hot vector. The experimental loop is inspired by \cite{Canatar_2021} and described in \cref{alg:experiment}, where the input conditions $X$ signify the datasets mentioned above, we create $\hat{c}^n_x$ with different setting of hyperparameters $n$ and $x$, and $\lambda = 0.0001$.

\begin{algorithm}
\caption{Experimental loop adapted from \cite{Canatar_2021}}\label{alg:experiment}
\begin{algorithmic}
\Require $X, \hat{c}, \lambda$
\State $K \gets kernel(X, X)$
\State  $\gamma, V_{\gamma} \gets eig(K)$ \Comment{Sorted according to $\gamma$}
\State  $\tilde{y} \gets$ \cref{eq:relabeling} $(V_{\gamma}, \hat{c})$
\While{$n < size(X)$}
    \While{$i < 50$}
        \State $K_n \gets kernel(X_n, X_n)$
        \State $\hat{y} \gets K_n \tilde{y}$
        \State $E_{empirical} \gets \frac{1}{n}\sum_i(\tilde{y}_i - \hat{y}_i)^2$
    \EndWhile
    \State $E_{predicted} \gets$ \cref{eq:generalization}$(\lambda, K, \tilde{y})$
\EndWhile

\end{algorithmic}
\end{algorithm}

As an additional illustration, we generate a semi-synthetic dataset tailored to both quantum $\hat{y}_Q$ and classical $\hat{y}_C$ KMs. To enable analysis beyond the discrete data points, we interpolate the data using SciPy's \texttt{RBFInterpolator}. The resulting multi-dimensional function is then plotted. Since visualizing an 8D function directly is not feasible, we create a 2D slice by selecting a random direction in 8D, constructing a random orthogonal direction, and moving along these two directions to sample points in a 2D subspace. We plot both $\hat{y}_Q$ and $\hat{y}_C$ on these slices (kept fixed between quantum and classical functions). This procedure is repeated three times to generate three random slices. We evaluate the smoothness of $\hat{y}_Q$ and $\hat{y}_C$ using the metrics defined in \cref{eq:lipschitz} and \cref{eq:sobolev} (for $k=2$) 
, and compute the correlations between these smoothness metrics and the corresponding $\hat{y}_Q$ and $\hat{y}_C$.

%% file: sections/5_results.tex
\subsection{Artificial labels}
In this section, we present and analyse the results of QKMs on the datasets we created with varying sizes. Our experiment aims to identify classical datasets where quantum learners achieve lower error rates with fewer data points than their classical counterparts. Additionally, we assess the alignment of theoretical predictions from \cref{eq:generalization} with empirical results. 

\begin{figure*}
\begin{subfigure}[t]{0.45\textwidth}
    \includegraphics[width=\textwidth]{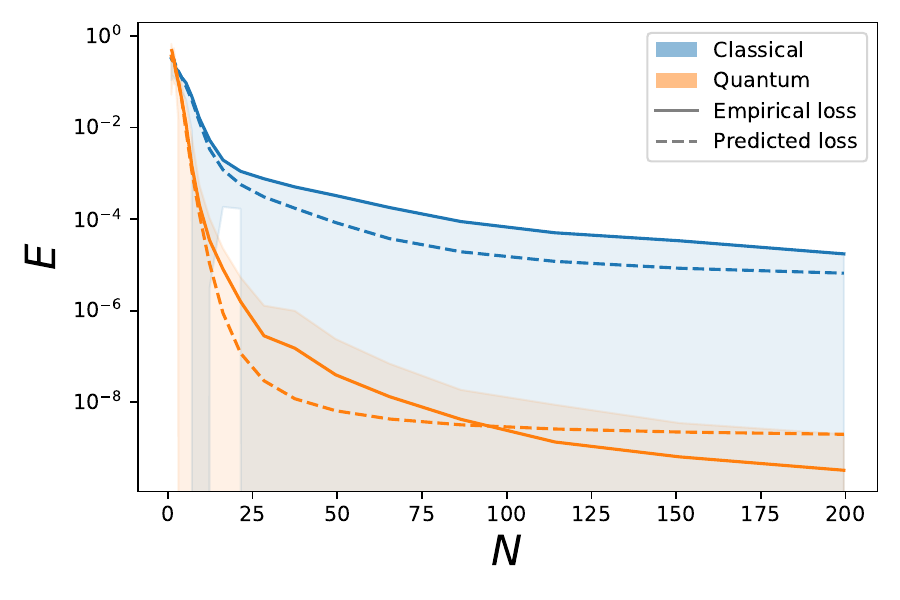}
    \caption{$\hat{c}^{4}_1$}
    \label{fig:catanar_4}
\end{subfigure}
\begin{subfigure}[t]{0.45\textwidth}
    \includegraphics[width=\textwidth]{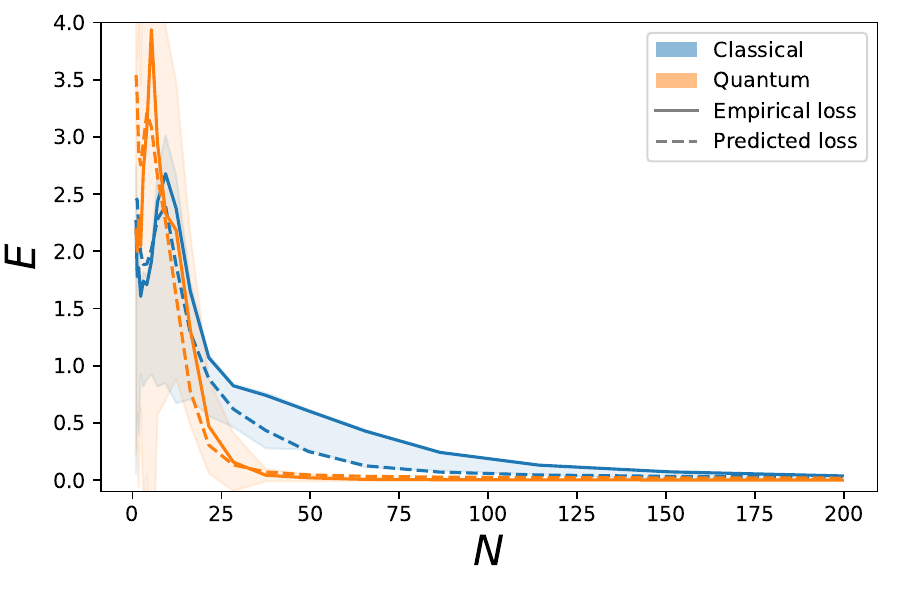}
    \caption{$\hat{c}^{20}_1$}
    \label{fig:catanar_20}
\end{subfigure}

\begin{subfigure}[t]{0.45\textwidth}
    \includegraphics[width=\textwidth]{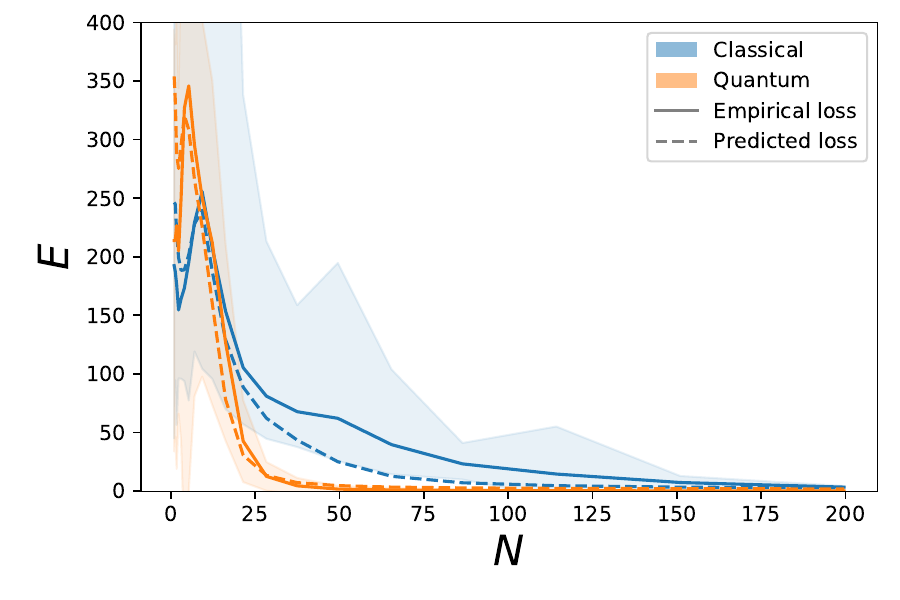}
    \caption{$\hat{c}^{20}_{10}$}
    \label{fig:catanar_20_10}
\end{subfigure}
\begin{subfigure}[t]{0.45\textwidth}
    \includegraphics[width=\textwidth]{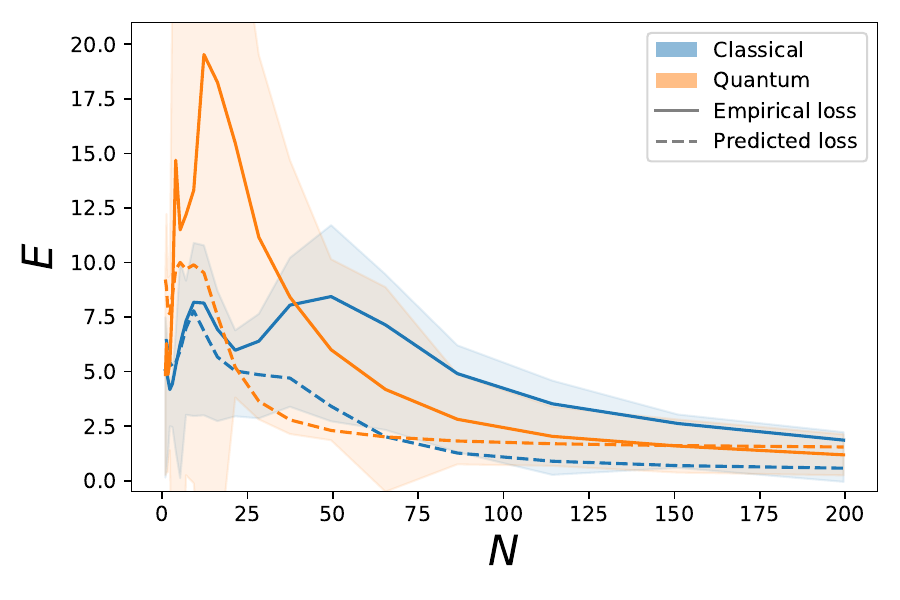}
    \label{fig:catanar_50}
    \caption{$\hat{c}^{50}_1$}
\end{subfigure}
\caption{Square mean error loss $E$ difference between classical and quantum kernels that depends on $N$ on dataset relabeled to saturate target alignment}
\label{fig:catanar_results}
\end{figure*}

As illustrated in \cref{fig:catanar_results}, we analyse the performance on relabeled datasets that saturates target alignment. The graphs demonstrate that, on average, the QKM achieves low error rates even with a small number of datapoints. By varying $\hat{c}^n_x$ we can create datasets of varying complexity, allowing for a nuanced analysis of quantum models' preference. The plots depicting different alignment strengths in \cref{fig:catanar_4}, \cref{fig:catanar_20} and \cref{fig:catanar_50} reveal that the convergence rates for QKMs differ significantly. Although this procedure did not explicitly account for a classical kernel (RBF), it is surprising to observe the disparity in the number of data points required compared to QKMs to achieve comparable error rates. To our knowledge this is the first empirical evidence that this is in fact possible. 

Hyperparameter $n$ seems to play a significant role in the relationship between QKMs and RBFs, as $n=4$ leads to only a small difference, while $n=50$ makes the curves more volatile. Hyperparameter $x$, on the other hand, seems to be controlling the scale, e.g. \cref{fig:catanar_20} and \cref{fig:catanar_20_10}. 

The overall trends of predicted loss from \cref{eq:generalization} remain consistent with empirical behavior. These results are promising, particularly given that other theoretical metrics for model generalization often exhibit low correlation with empirical accuracy in previous studies \cite{Egginger_2024, monnet2024understandingeffectsdataencoding}. This indicates that generalization error from \cref{eq:generalization} is a promising candidate for further studies of generalization in QML. For completeness, we investigated an alternative method for generating artificial labels in \cref{sec:geometric_diff}.

\subsection{Labeling function complexity analysis}
In this section we present a preliminary example of how the artificial labels introduced above can illuminate use cases where quantum kernel methods (QKMs) may offer advantages. 

\begin{figure}
\centering
    \includegraphics[width=0.5\textwidth]{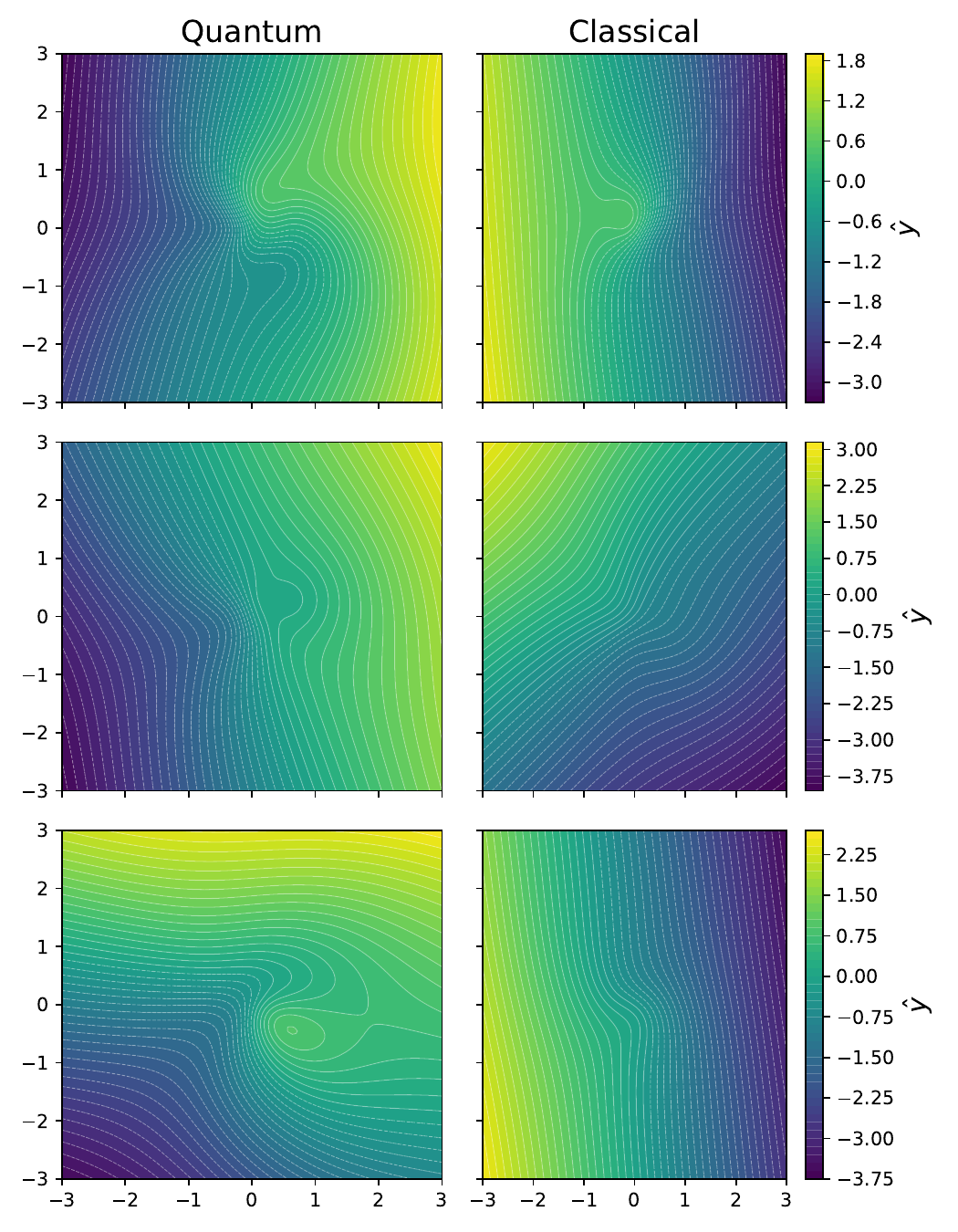}
\caption{Function landscape of the labeling function generated for quantum versus classical KMs (columns). Each row represents a randomly chosen 2D slice (randomly chosen direction and its orthogonal) of the multidimensional function.}
\label{fig:dataset_landscape}
\end{figure}

\cref{fig:dataset_landscape} shows a visualization of landscapes of labeling function generated for quantum $\hat{y}_Q$ and classical $\hat{y}_C$ KMs. The 2D slices shown are selected at random from among the rows of the figure and are held fixed across $\hat{y}_Q$ and $\hat{y}_C$ to enable a fair comparison. Because the slice selection is random, the results should be interpreted with caution; nonetheless, some tendencies emerge. To make these trends easier to discern, the range of probed points is chosen slightly larger than in the original dataset. The plots suggest that the quantum kernels favor functions with certain curvature, whereas the classical RBF kernel tends to produce smoother, more gradual changes.

\begin{figure}
    \centering
    \includegraphics[width=0.9\linewidth]{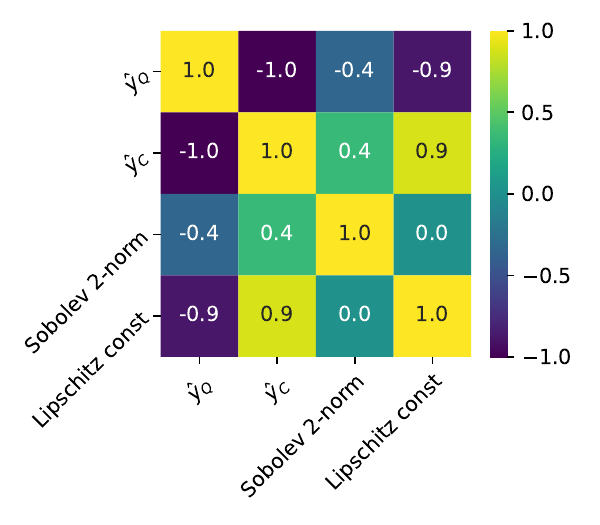}
    \caption{Correlation heatmap between labeling function generated for quantum $\hat{y}_Q$ versus classical $\hat{y}_C$ KMs and different smoothness metrics (\cref{eq:lipschitz} and \cref{eq:sobolev}).}
    \label{fig:dataset_correlation}
\end{figure}

\cref{fig:dataset_correlation} shows a correlation heatmap among $\hat{y}_Q$, $\hat{y}_C$ and different smoothness metrics defined in \cref{eq:lipschitz} and \cref{eq:sobolev}. Notably, $\hat{y}_Q$ exhibits a strong negative correlation with the Lipschitz constant $L$, indicating a tendency toward Lipschitz-smooth functions. In contrast, $\hat{y}_C$ tends to have larger $L$ values, yielding a strong positive correlation with $L$. Because the $L$ values for both label types lie in a small to moderate range, this strong positive correlation should not be interpreted as a general preference of classical RBF kernels for rapidly varying functions. Another observation is that the Sobolev norm with k=2 shows a weak negative correlation, contrary to what is suggested by the landscapes in \cref{fig:dataset_landscape}. One of the explanations for this behavior is that this norm does not solely depend on the 2-degree (which captures the curvature) but also incorporates the 1-degree derivative. Since the there is a strong negative correlation with the Lipschitz constant, which is based on 1-degree derivatives (for differentiable functions), it could have offset the curvature effects in this metric. This indicates that additional metrics, e.g. based on L2-norm of a Hessian matrix, should be explored to characterized these visual artifacts better. 

Overall, while not conclusive on their own, these results demonstrate the usefulness of our artificial labeling tool. It enables the generation of semi-artificial datasets, which characteristics can be studied. This approach can guide the search for real-world datasets that exhibit the identified characteristics, as an alternative to relying on the currently popular practice of randomly sampling real data in the hope of finding instances where QML offers an advantage.

%% file: sections/6_discussion.tex
There are various speculations in the literature regarding the possibility of data-efficient learning in QML \cite{kubler2021inductive, Huang_2021, Jerbi_2023}, particularly when using classical data. This uncertainty served as a significant motivation for our work, as we aimed to investigate whether data-efficiency is achievable with quantum kernels compared to classical kernels. Another motivating factor for this study is the promising connection between Quantum Kernel Machines (QKMs) and Deep Learning Models (DLMs) through Neural Tangent Kernels. This relationship presents an intriguing avenue for future research to explore the implications of our findings across a broader range of models. Notably, the work by \cite{Jerbi_2023} suggests that QKMs may require exponential amounts of data, in contrast to models that are more akin to quantum versions of DLMs. This could imply that the insights derived from our proposed tools may serve as an upper bound for other quantum models that require even less data.

Our study highlights an important tool for understanding which types of datasets are best suited for QKMs. We demonstrated a simple usage scenario of the tool: we generated datasets for classical and quantum KMs and examined their properties. A visual inspection of the resulting landscapes reveals intriguing functional artifacts, but the chosen metrics do not fully capture them (e.g. Sobolev 2-norm did not show positive correlation). In future work, we will adopt more robust measures of functional complexity and scale the experiments to reveal the hidden inductive biases that distinguish quantum from classical KMs. Another important avenue for future research is to modify features of the dataset (beyond just the labeling function as done in this work), which opens up a wider range of potential dataset complexities analysis. These complexities may stem from specific types of noise \cite{schetakis2021binaryclassifiersnoisydatasets}, geometric properties such as symmetry \cite{Meyer_2023}, or the evaluation of datasets characterized by high intrinsic dimensionality \cite{Huang_2021}. Investigating these dataset properties could ultimately help us identify real use cases where quantum models offer a distinct advantage.

The goal of the method we proposed in this work was not to maximize the difference between classical and quantum learners but rather to develop a dataset that is more favorable to quantum learners. An intriguing future direction is to integrate concepts from \cite{Huang_2021} to develop strategies that maximize the performance gap between classical and quantum methods. This approach could lead to a more controlled quantum-classical gap and increased diversity in the datasets. Furthermore, scaling this approach to identify data-efficiency gaps that are of industrial interest represents another promising research avenue.

A surprising finding from our study is that changing the target function alone can lead to significant changes in performance. Notably, our method did not alter the features of the dataset, and therefore had no impact on the eigenvalues. An intriguing direction for further research is to explore how dataset features and the architectural properties of quantum models affect eigenvalues of the resulting kernel matrix, potentially including a theoretical derivation of eigenvalue requirements. Such properties of models can include e.g. bandwidth. Inspired by the study \cite{canatar2023bandwidthenablesgeneralizationquantum} on the importance of bandwidth, we conducted a supplementary analysis presented in \cref{sec:bandwidth}; however, further investigation is warranted.